\documentclass[prd,aps,preprint,amsmath,nofootinbib,amssymb,eqsecnum,showkeys,tightenlines]{revtex4-1}
\usepackage{slashed}
\usepackage{epsfig,latexsym,cancel,amssymb,amsmath,verbatim,mathrsfs}
\usepackage{color}
\usepackage{graphicx}

\def\ra{\rightarrow}
\def\L{\left(}
\def\R{\right)}
\def\wt{\widetilde}
\def\Ld{\Lambda}

\def\f{\frac}
\newcommand{\be}{\begin{equation}}
\newcommand{\ee}{\end{equation}}
\newcommand{\bea}{\begin{eqnarray}}
\newcommand{\eea}{\end{eqnarray}}
\newcommand{\ba}{\begin{array}}
\newcommand{\ea}{\end{array}}

\long\def\symbolfootnote[#1]#2{\begingroup%
\def\thefootnote{\fnsymbol{footnote}}\footnote[#1]{#2}\endgroup}

\newcommand{\beq}{\begin{equation}}
\newcommand{\eeq}{\end{equation}}

\begin{document}

\title{Slightly Ultra-violet  Freeze-in a Hidden Gluonic Sector}

\author{Zhaofeng Kang}
\email[E-mail: ]{zhaofengkang@gmail.com}
\affiliation{School of physics, Huazhong University of Science and Technology, Wuhan 430074, China}

\date{\today}

\begin{abstract}

The dark glueball (DGB) from a hidden Yang-Mills sector is a simple non-WIMP dark matter candidate characterized by very few parameters. However, it suffers the over dense issue. To overcome it, in general the dark sector is required to be hierarchically cooler than the visible sector. To naturally generate the desired hierarchy, in this paper we introduce higher dimensional operators  coupling the dark gauge field strength tensor to the standard model (SM) Higgs doublets or gauge field strength tensors. By tracking the different phases of the universe from the end of inflation, prethermalization, reheating to the radiation dominance era, we show that these operators can make the DGB be a viable dark matter candidate over a wide mass region, from the sub-GeV to multi-PeV or even beyond. At the same time, these operators open decay channels of DGB to the SM species, and partial of the parameter space could leave hints in the cosmic ray.

\end{abstract}


\maketitle

\section{Introduction}

It becomes more and more channeling to naturally reconcile the weakly interacting massive particle (WIMP) paradigm of dark matter (DM) and a bunch of strong exclusions limits on the strength of DM-SM (standard model) interactions. The dark glueballs (DGBs), predicted by a hidden $SU(N_d)$ pure Yang-Mills theory in the confining phase, may furnish one of the simplest non-WIMP DM candidate; its nature is specified by very few parameters. Not a WIMP, it can easily resolve the puzzle of null DM searches. In addition to that, the $SU(N_d)$ non-Abelian gauge sector is well expected in various contexts of new physics beyond the SM such as dark matter/radiation~\cite{DM1,DM2,Ko:2017uyb}, origin of baryon asymmetry~\cite{baryon}, naturalness problem~\cite{twin}, and especially in the string theory~\cite{string}.

One attractive merit of the DGB DM is that its mass is generated dynamically. There is a tower of DGBs classified by their quantum numbers labeled as $J^{PC}$, and they are absolutely stable without a connection to the SM. The one lying at the bottom, namely $J^{PC}=0^{++}$, is supposed to be the dominant DM components. We will concentrate on this state throughout this paper. The mass of DGBs are determined by the confinement scale $\Ld_d$. According to the lattice simulation without introducing the topological $\theta$ term~\cite{lattice}, the mass of $0^{++}$ is 
\begin{eqnarray}\label{d=5}
m_0=\L\alpha+\beta/N_d^2\R \Ld_d,
\end{eqnarray}
where $\alpha$ and $\beta$ are order one parameters. In the large $N_d$ limit, the correction from color number is suppressed by $1/N^2_d$ and therefore the mass approaches a constant universal to all $SU(N_d)$. From the well-studied $N_d=3$ case, $\alpha\approx 7$~\cite{lattice}. One may expect that the mass spectra of $N_d>3$ assembles that of $SU(3)$.

How to get the correct relic density of the dark glueball DM is an issue. Usually, it is assumed that the dark gluons form a thermal bath at the early universe, experiencing the highest temperature ${\cal T}_{max}$ far above the confining scale $\Ld_d$. When ${\cal T}$ cools down below $\Ld_d$, the confining phase transition happens and the dark gluons turn into the non-relativistic dark glueballs. In terms of the effective theory for $0^{++}$, its number density freezes-out after the $3\ra 2$ process incapable of catching the Hubble expansion rate. It is found that the DGB overcloses the universe, except for ${\cal T}(\Ld_d)\ll T(\Ld_d)$, namely a much cooler dark sector than the SM sector. The above conclusion holds no matter whether the two sectors used to reach thermal equilibrium or not.

In order to produce DGBs in the early universe, we should introduce proper interactions for the $SU(N_d)$ dark sector. A popular way to link it with the SM sector is adding higher dimensional operators in the form of ${\cal O}_{\rm SM}{\cal O}_{d}$, suppressed by very high cut-off scale in order to guarantee a sufficiently long lifetime of $0^{++}$. In this setup, the dark gluons can be slowly produced through the SM species scatterings with rates which are very suppressed by the cut-off scale appearing in the higher dimensional operators. Consequently the dark gluons fail to keep in thermal equilibrium with the SM sector, and thereby, as desired, they naturally have a much lower temperature than the SM bath. This way of producing the dark sector is nothing but an application of the UV freeze-in mechanism~\cite{Hall:2009bx,Elahi:2014fsa}. Different than the usual freeze-in via renormalizable operators which is not senstive to UV~\cite{McDonald:2001vt,Hall:2009bx,Bernal:2017kxu,Kang:2014cia}, in UV freeze-in special attention should be paid to the very early universe before reheat, when the universe went through eras with an even higher temperature than the reheat temperature~\cite{Co:2015pka,McDonald:2015ljz,Garcia:2017tuj,Chen:2017kvz,Garcia:2018wtq,Forestell:2018dnu}.

The goal of this paper is to detailedly investigate if the $d=6$ and $d=8$ operators could appropriately reheat the dark gluonic sector via the UV freeze-in mechanism. We take into account the contributions from different phases of the universe: 1) Phase I just after inflation during which the radiation from inflaton decay has not reached thermalizaiton yet; DM production in this phase receives attention just very recently~\cite{Garcia:2018wtq}. 2) Phase II after thermalization of radiation but before the completion of reheat. 3) Phase III, the usual radiation era after reheat. We find that the contribution from phase I can be the dominant one in the $d=8$ case, provided that a very heavy inflaton decaying slowly, for instance via Planck suppressed operators. Whereas in the $d=6$ case dark gluons tend to be thermally produced.

The paper is organized as the following: In Section II we present the effective model of DGB and introduce higher dimensional  operators bridging the dark and visible sectors, and we also present the resulting decay widths of DGB. In Section III we focus on how the higher dimensional operators slightly reheat the dark sector. Section IV contains the conclusions and discussions.

\section{A decaying dark glueball}

\subsection{The self-interaction \& relic density issues of DGB dark matter}

At the energy scale below the scale $\Ld_d$ where the $SU(N_d)$  gauge coupling becomes strong, the degrees of freedom of the theory are DGBs instead of gluons. But how to construct the low energy effective theory for these new ingredients is  out  of the comfort zone of QFT.  The following effective model for the lowest lying state $0^{++}\equiv s$~\cite{Forestell:2017wov} is proposed in the spirit of  non-perturbative methods,
\begin{align}\label{EFT}
{\cal L}_{\rm DBG}= \f{1}{2}\L\partial s\R^2-\f{1}{2!}m_{s}^2s^2-\f{c_3}{3!}\f{4\pi}{N_d}m_{s} s^3-\f{c_4}{4!}\L\f{4\pi}{N_d}\R^2 s^4-\f{c_5}{5!}\L\f{4\pi}{N_d}\R^3\f{s^5}{m_{s}} -...
\end{align}
where $c_i$ are expected to be order one numbers in the light of the large-$N$ limit and naive dimension assumption (NDA). However, it is definitely not  beyond dispute. Different viewpoints lead to significant differences in the estimation of couplings. In Ref.~\cite{Boddy:2014yra} the estimation is just based on NDA without considering the large-$N$ limit, then the $1/N_d$ suppression factor is absent. On the contrary, in Ref.~\cite{Soni:2016gzf} which just follows the large-$N$ limit, the $4\pi$ factor is not presented. Thus, we may keep an open mind on the strength of these couplings. In this paper we follow Eq.~(\ref{EFT}) unless explicitly stated otherwise. The values of effective couplings have direct implications to the self-interaction and relic density of the dark matter candidate $s$.

First, the self-interacting term $s^4$ gives rise to the elastic scattering $ss\to ss$, which has a large cross section $\sigma_{2\ra 2}\sim (4\pi/N_d)^4/m_{s}^2$. The Bullet Cluster imposes an upper bound on  $\sigma_{2\ra 2}/m_{s}  \lesssim 1{\rm cm}^2/g$~\cite{bullet}, which in turn gives the lower bound on the DGB mass~\cite{DGB,Forestell:2017wov},
\begin{eqnarray}\label{msbound}
m_{s}\gtrsim 100\times (3/N_d)^{4/3}{\rm MeV}.
\end{eqnarray}
 In the large-$N$ limit, DGB is supposed to be weakly interacting and thus could relieve this bound, but not much given a normally large $N_d$.  In addition to leave hints in the large scale structure today,  this fast elastic scatterings $ss\to ss$ could keep the DGBs in thermal equilibrium within themselves just after the dark confining phase transition.

Second, the $s^5$ term leads to a sizable particle number depletion process $3s\ra 2s$, which could maintain the chemical equilibrium  among the DGBs. After the $3s\ra 2s$ depletion rate falls below the Hubble expansion rate  at ${\cal T}_d$, the chemical equilibrium is lost and the  DGB number density  freezes out.  As a consequence of entropy conservation in both sectors, the relic density is determined to be~\cite{DGB,DGB:relic,Forestell:2017wov},
\begin{eqnarray}\label{relic0}
\Omega_s h^2\simeq \f{g^{d}_S\xi_T^3}{g_S}\f{{\cal T}_d}{3.6\rm eV},
\end{eqnarray}
where $g_S$ and $g_S^d$ are the entropy degrees of freedom in the visible and dark sectors densities, respectively; $\xi_T\equiv {\cal T}/T$ is the ratio between the temperatures of the two sectors. All of them are calculated at a temperature ${\cal T}_i\gtrsim \Ld_d$, as the initial conditions for the DGB freeze-out era. And they are constant up to any high temperature until entropy conservation is violated. 

It is expected that ${\cal T}_d\lesssim \Ld_d$. In particular, ${\cal T}_d\simeq \Ld_d$ in the limit $c_5\ra 0$~\cite{Boddy:2014yra,Soni:2017nlm}. Otherwise, one may parameterize the deviation by ${\cal T}_d=w_f\Ld_d$ with $w_f\lesssim 1$. Anyway, from Eq.~(\ref{relic0}) we immediately see that, as stated in the introduction, $\xi_T\ll1$ is indispensable to reduce the large ratio ${\cal T}_d/(3.6 {\rm eV})\simeq \Ld_d/(3.6 {\rm eV})\gtrsim 10^8$ indicated by Eq.~(\ref{msbound}). The main goal of this paper is to specify the UV freeze-in as the natural mechanism for generating  $\xi_T\ll 1$  .

\subsection{Higher dimensional operators and the decaying dark glueball}

Although it is possible that the dark $SU(N_d)$ Yang-Mills sector thus dark glueballs interact with the SM sector purely gravitationally, many people introduce messengers communicating interactions between the two sectors. These messengers are supposed to be charged under $SU(N_d)$ and at the same time communicating with the SM particles via gauge or Yukawa gauge interactions. Moreover, they are very heavy, having masses $m_Q\gg \Ld_d$. Alternatively, the messengers may be a heavy moduli-like field which is neutral under any gauge groups~\cite{Chowdhury:2018tzw}. After integrating out the messengers one obtains the higher dimensional operators ${\cal O}_{\rm SM}{\cal O}_{d}$ that describe the interactions between the dark glueball and SM species. The concrete effective Lagrangian is model dependent, on the choice of messengers. In this article, to demonstrate the idea, we just consider two representative examples at the $d=6$ and $d=8$ level, respectively,
\begin{eqnarray}\label{EFT}
{\cal L}_{int}=\f{g_d^2}{M^2_6}{\rm tr}{\cal F}^{\mu\nu}
{\cal F}_{\mu\nu}|H|^2 +
\f{g^2_d}{M^4_8}{\rm tr}{\cal F}^{\mu\nu}
{\cal F}_{\mu\nu}{\rm tr}{G}^{\mu\nu}
{G}_{\mu\nu}
\end{eqnarray}
where $M_{6,8}$ are the cut-off scales, which in a concrete UV model are the combinations of the heavy messenger mass scales and various couplings, but for later use here we leave $g_d^2$ with $g_d$ the dark QCD gauge coupling. The dark QCD gluon field strength tensor is written as ${\cal F}_{\mu\nu}={\cal F}^a_{\mu\nu}T^a$, with $T^a$ the generators of $SU(N_d)$ satisfying the  normalization conditions ${\rm tr}(T^aT^b)=\delta^{ab}/2$. The same convention applies to other non-Abelian gauge groups.  We only consider the scalar operator $S\equiv {\rm tr}{\cal F}^{\mu\nu}
{\cal F}_{\mu\nu}$, and likely other operators accompany, e.g., ${\rm tr}{\cal F}^{\mu\nu}
\wt{\cal F}_{\mu\nu}{\rm tr}{G}^{\mu\nu}\wt
{G}_{\mu\nu}$ and ${\rm tr}{\cal F}^{\mu\alpha}
{\cal F}_{\alpha}^\nu {\rm tr}{G}_\mu^\alpha
{G}_{\alpha\nu}$, and so on~\cite{DG:eff1,DG:eff2}. But their presence does not change the main line of our discussion, though a possible sizable numerical correction~\cite{Ellis:2018cos}.

Several comments on UV models are in orders. First, the Higgs-portal operator selects a class of UV complete models, for instance those with scalar messengers which are neutral under the SM gauge groups but have Higgs portal interactions~\cite{Cline:2013zca}. Otherwise, one may have to arrange the proper Yukawa interactions between the fermionic messengers and the Higgs doublet; see Ref.~\cite{DG:eff2}. Second, at the $d=8$ level if the messengers are color neutral, the portal does not contain gluons; on the contrary, the portal might be pure gluon-portal provided that the messengers do not carry electroweak quantum numbers. In the context of grand unification, it is expected that all kinds of SM vector bosons are present. Third, $M_{6,8}$ do not directly correspond to the mass scale of loop particles in the UV model, but usually the difference is not big if there are no very small/large extra Yukawa couplings involved in the loop.

The operators ${\cal O}_{\rm SM}{\cal O}_{d}$ may open decay channels for the DGBs into the SM sector. The calculation of their decay width is straightforward given the factorized matrix elements, which for a certain DGB state $J^{CP}$ decaying into the SM states without involving another DGB in the final state is~\footnote{Transitions between dark blueballs are at the radiative level and thus may play significant roles in the models where the DGBs decay very fast without protection by quantum numbers. But in our scope, where $J^{CP}=0^{++}$ is the dominant DM component, the leading order consideration is sufficient.}
\begin{eqnarray}\label{}
 \langle {\rm SM}|{\cal O}_{\rm SM}|0\rangle\langle 0|{\cal O}_{d}|J^{CP}\rangle.
\end{eqnarray}
In this matrix element, the SM part can be calculated perturbatively, but the other part must fall back on non-perturbative methods and is parameterized as $F_{J^{CP}}^{{\cal O}_d}$, known as the decay constant of the state $J^{CP}$. Hereafter we will specify $J^{CP}=0^{++}\equiv {s}$ and take $F_{0^{++}}^{S}$ as $F_{{s}}$. From the lattice result~\cite{FS}, one may take
\begin{eqnarray}\label{FS}
{g_d^2} F_{s}\simeq {3.06}m_0^3.
 \end{eqnarray}
For a given ${\cal O}_d$, the accessible annihilation decay modes of ${s}$ depend on its mass. In the following we will discuss the decay pattern of DGB separately in the $d=6$ and $d=8$ cases.

\subsection{Higgs-portal DGB ($d=6$)}

For the Higgs portal case, the decay pattern of ${s}$ is well-understood as an additional Higgs state mixing with the SM Higgs boson, which for a wide mass region of $m_h$ has been widely studied~\cite{Hdecay}. The partial decay width of ${s}$ into a pair of SM states is written as
\begin{eqnarray}\label{}
\Gamma_{{s}\ra {\rm SM}+{\rm SM}}=\L\f{vg_d^2F^{s}}{ M^2(m_s^2-m_h^2)}\R^2\Gamma(h\ra {\rm SM}+{\rm SM})|_{m_h\ra m_s},
\end{eqnarray}
where the factor in the bracket could be identified as the mixing angle between $h$ and ${s}$; $v=246$ GeV and $m_h=125$ GeV is the SM-like Higgs mass. If the DGB is heavy having a mass $m_{s}=m_0\gg m_W$, then $s$ dominantly decays into the vector bosons and the resulting lifetime of $s$ is estimated to be
\begin{eqnarray}\label{SVV}
\tau_{s}\simeq 1.1\times10^{30} \L\f{M_6}{ 10^{17}\rm  GeV}\R^4\L\f{100\rm GeV}{\Ld_d}\R^5 s.
\end{eqnarray}
In order to evade the constraints on decaying DM into vector bosons~\cite{Cohen:2016uyg}, most stringently by the diffuse gamma ray searches, the lifetime must be much longer than the age of the universe. For the decaying DGB DM around the TeV scale, it suggests that the interactions between two sectors actually must be suppressed by the near-Planck scale.

In the opposite limit, the DGB is very light and lies much below the weak scale but above 100 MeV due to the self-interaction bound Eq.~(\ref{msbound}). Then, ${s}$ dominantly decays into the heaviest SM fermion pair $f\bar f$ kinematically accessible. Probably the most suppressed case of DGB decay is that it dominantly decays into a pair of muons for $1{\rm GeV}\gtrsim m_s\gg 2m_\mu$. In this case one has
\begin{eqnarray}\label{Smu}
\tau_{s}\simeq 5.3\times10^{27} \L\f{M_6}{ 10^{10}\rm  GeV}\R^4\L\f{0.1\rm GeV}{\Ld_d}\R^7\L\f{0.1\rm GeV}{m_f}\R^2 s.
\end{eqnarray}
Even if the region $2m_e\ll m_{s}\lesssim 2m_\mu$ is marginally realized after taking into account the uncertainty of the lower bound on $m_{s}$, the above estimation will not be dramatically changed. However, if we do not insist on this bound and allow a much lighter DGB then the scale of $M_6$ can be fairly low. Now since $m_{s}\ll 2m_e$, only a pair of photon is accessible in $s$ decay and the resulting DGB lifetime is 
 \begin{eqnarray}\label{}
\tau_{s}\simeq 2.3\times10^{26} \L\f{M_6}{ 10^{4}\rm  GeV}\R^4\L\f{1\rm MeV}{\Ld_d}\R^9 s.
\end{eqnarray}
It is seen that in this case the cutoff scale can be as low as the interesting TeV scale. Therefore, the lower bound on $\Ld_d$ by virtue of self-interaction places a strong lower bound on $M_6$.


\subsection{Vector-boson-portal DGB ($d=8$)}

Now we move to the next case, DGBs communicating with the SM sector via the vector-boson-portal at the dimension-8 level. $s$ can annihilate decay into a pair of vector bosons such as a pair of free QCD gluons, and the decay width is~\cite{DG:eff2}
\begin{eqnarray}\label{sgg1}
\Gamma_{s\ra gg}=\f{1}{2\pi} \f{1}{M_8^8} m_{s}^3(g_d^2F_s)^2.
\end{eqnarray}
Compared to the $d=6$ case, it has a much stronger dependence on the confining scale $\Ld_d$, which then leads to a much longer lifetime of DGB. Note that if the DGB mass is below the GeV scale, the hadronic mode $S\ra gg$ is closed today. The lifetime of the decaying DGB in terms of Eq.~(\ref{sgg1}) is estimated to be
\begin{eqnarray}\label{sgg2}
\tau_{s}\simeq 1.1\times10^{27} \L\f{M_8}{ 10^{8.5}\rm  GeV}\R^8\L\f{10\rm GeV}{\Ld_d}\R^9  s.
\end{eqnarray}
Lowering $\Ld_d$ down to the sub-GeV scale, $M_8$ can be down to the PeV scale. In our analysis, we focus on the DGB mass regions which admit a simple analytical expressions of decay width,that is convenient to our final numerical display.

\section{Higher dimensional operators: Reheat the dark sector}

The UV freeze-in mechanism is sensitive to the scattering process characterized by hard momentum, so we have to backdate the universe to the extremely hot state just after inflation. It is usually known as the reheating stage and gives the broadly quoted ``highest" temperature of the universe $T_{re}\sim \sqrt{\Gamma_\phi M_{\rm Pl}}$, with $\Gamma_\phi$ the width of the inflaton perturbative decay to radiation which thermalizes instantaneously at $t_{re}\simeq 1/\Gamma_\phi$. But there is an even hotter phase than $T_{re}$ during reheating~\cite{reheating}, whose possible impacts on UV freeze-in are investigated by several groups~\cite{Co:2015pka,McDonald:2015ljz,Chen:2017kvz}, to find that the impacts are small given $d<9$. These studies assume that the radiation is always instantaneously thermalized during reheating. However, instantaneous thermalization may be a bad approximation. Actually, during reheating there exists a phase prior to the thermal equilibrium of radiation, namely prethermalization when the radiation has not reached full thermal equilibrium yet. Radiation in this phase is dominated by the extremely hard primary modes with momentum around the inflaton mass $m_\phi$, and thus the UV freeze-in process, whose cross section is greatly enhanced, may be very effective in this phase. Ref.~\cite{Garcia:2018wtq} carefully studied the production of FIMP DM during the prethermlization era and found that it indeed could be the dominant contribution to the final FIMP relic density. 


Before proceeding, we briefly review the thermalization process studied in Ref.~\cite{Harigaya:2013vwa}. The inflaton decay width is parameterized as $\Gamma_\phi=k m_\phi^2/M_{\rm Pl}$, from which one can see that $k\gtrsim 1$ leads to $T_{re}\gtrsim m_\phi$ and thus thermal effects on the inflaton decay become important~\cite{thermal,k>1}. To avoid this complication, in this paper we just focus on the slow decay limit with $k\ll1$, for instance, inflaton decay via a Plank suppressed dimension five operator. Furthermore, it is reasonable to consider that  the radiation is charged under a non-Abelian gauge group with normal gauge coupling $\alpha\sim{\cal O}(0.01)$. The energy of primary hard radiation could be effectively dissipated away via the fast collinear soft gauge boson emissions, eventually reaching thermalization. It is shown that as long as $\Gamma_\phi\ll \alpha^8 M_{\rm Pl}^3/m_\phi^2$, or equivalently,
\begin{eqnarray}\label{lower}
k\L m_\phi/M_{\rm Pl}\R^{4}\ll \alpha^{8},
\end{eqnarray}
thermalization is completed at a time scale $t_{th}$ well before the reheating time scale $t_{re}$. Concretely, the thermalization time scale is determined by
\begin{eqnarray}\label{thermalization}
t_{th}-t_{end}\simeq\Gamma^{-1}_\phi\alpha^{-16/5}\L\f{\Gamma_\phi m_\phi^2}{M_{\rm Pl}^3}\R^{2/5},
\end{eqnarray}
with $t_{end}$ denoting the end of inflation.

In what follows, we will follow the thermal history of the universe, to describe how UV freeze-in via the higher dimensional operators just provides a good way to slightly reheat the dark gluonic sector.

\subsection{UV freeze-in dark gluons}

\subsubsection{The BEs for the inflaton-radiation-dark gluon system}

The dark gluons are radiation, so it is more useful to derive BE for the energy density instead of number density. During the reheating region, the BEs for the inflaton-radiation-dark gluon system are
\begin{eqnarray}\label{inflaton}
\f{d}{dt}\rho_\phi+3H\rho_\phi+ \Gamma_\phi \rho_\phi&  = & 0,\\\label{radiation}
\f{d}{dt}\rho_R+4H\rho_R-\Gamma_\phi \rho_\phi+\wt\gamma_{g_d} &= &0.\\
\label{darkG}
\f{d}{dt}\rho_{g_d}+4H\rho_{g_d}-\wt\gamma_{g_d}&=& 0,
\end{eqnarray}
where the collision term $\wt \gamma_d$ will be specified later. The Hubble parameter is determined by the Friedmann equation:
\begin{eqnarray}\label{frie}
H\equiv\dot a/a=\sqrt{(\rho_\phi+\rho_R+\rho_{g_d})/3M_{\rm Pl}^2}.
\end{eqnarray}
In this set of BE describing the energy density transfer,  inflaton dominates the system throughout the reheating stage thus $H\approx \sqrt{\rho_\phi/3M_{\rm Pl}^2}$. Its energy density is simply red shifting, and therefore $H \propto a^{-3/2}$ or $H(a)=H_{end}\L a_{end}/a\R^{3/2}$, where $H_{end}= \sqrt{\rho_{end}/3M_{\rm Pl}^2}$ with $\rho_{end}$ the energy density of inflaton at the end of inflation. In a class of popular inflation models $\rho_{end}$ is estimated to be
\begin{eqnarray}\label{}
\rho_{end}\sim m^2_\phi M^2_{\rm Pl}, 
\end{eqnarray}
From the Friedmann equation one can estimate  the age of the universe $t=\int_{a_{end}}^{a(t)}da'/( a'H(a'))$, arriving at the $t-a-H$ relation
 \begin{eqnarray}\label{ta:re}
t=\f{3}{2}\L\L \f{a}{a_{end}}\R^{3/2}-1\R H^{-1}_{end}\approx\f{3}{2}H^{-1}.
\end{eqnarray}
The last relation holds for $a\gg a_{end}$. This $t-a$ scaling rule is different than the one in the radiation dominant era Eq.~(\ref{ta:rd}) because the reheating era is matter dominant.

Reheating is completed when the inflatons decay away at $t_{re}\simeq 1/\Gamma_\phi$, after which the inflaton domination gives way to a period of radiation domination. Radiation energy density contains two components, the ordinary $\rho_R$ produced by inflaton decay and the dark radiation  $\rho_{g_d}$ produced by the ordinary radiation. Such a system is described by Eq.~(\ref{radiation}) and Eq.~(\ref{darkG}) with $\Gamma_\phi\ra 0$.~\footnote{It is assumed that the transformation from phase II to phase III is prompt. A more accurate treatment should take into account the decaying term of inflaton.} During this era $\rho_R$ dominates over $\rho_{g_d}$ and $H\approx \sqrt{\rho_R/3M_{\rm Pl}^2}\propto a^{-2}$. Then similar to the previous derivation one gets the $t-a-H$ relation
  \begin{eqnarray}\label{ta:rd}
t=\f{1}{2}\L\L \f{a}{a_{re}}\R^{2}-1\R H^{-1}_{re}\approx\f{1}{2}H^{-1}.
\end{eqnarray}
It is ready to show that in any phase $dt/da=H^{-1}a^{-1}$.

Simplifications to the BEs can be made. To scale out the effect of the universe expansion, we consider evolution of the dimensionless quantities $\Phi=\rho_\phi a^3/m_\phi$, $R=\rho_R a^4$ and $G_d=\rho_{g_d} a^4$. We also introduce the dimensionless variable $\hat a\equiv a m_\phi$. The radiation~\footnote{It is could be the SM gluons or any other members which are charged under the SM non-Abelian gauge groups, such as the Higgs doublet.} is gradually built up from the inflaton perturbative decay $\phi\ra RR$, while the dark sector is only slightly reheated by the radiation scattering $R+R'\ra g_d+g_d'$. These considerations motivate the no back reaction approximation: In the inflaton dominance era,  the energy leaking to the radiation is negligible and thus in Eq.~(\ref{inflaton}) the $\Gamma_\phi\rho_\phi$ term can be dropped; similarly,  in Eq.~(\ref{radiation}) the  $\wt \gamma_{g_d}$ term is removed.  Then, the BEs take the simple form during reheating,
\begin{eqnarray}\label{inflaton}
\Phi'&  \approx &0,\\\label{}
R'&  \approx & \f{\sqrt{3}M_{\rm Pl}\Gamma_\phi}{m_\phi^2}\Phi^{1/2} \hat a^{3/2},\\
\label{}
G_d'&  = &\f{\sqrt{3}M_{\rm Pl}}{m_\phi^6} \wt\gamma_{g_d} \Phi^{-1/2}{\hat a}^{9/2}.
\end{eqnarray}
where the prime denotes the derivative with respect to ${\hat a}$. While during the radiation dominated era, the BEs are reduced to
\begin{eqnarray}\label{rd:rasol}
R'&  \approx 0& ,\\
\label{DG:re}
G_d'&  = &\f{\sqrt{3}M_{\rm Pl}}{m_\phi^6}\wt\gamma_{g_d} \hat a^{5}R^{-1/2}.
\end{eqnarray}

In the no back reaction limit, it is ready to solve the BEs one by one. First, the solution to Eq.~(\ref{inflaton}) is
\begin{eqnarray}\label{inflaton:sol}
\Phi(\hat a)\approx \f{4}{3}\f{M_{\rm Pl}^2}{m^4_\phi}\hat a_{end}^3H_{end}^2,
\end{eqnarray}
which is a constant as expected, because the inflaton energy density is (approximately) purely red-shifting as $a^{-3}$. To account for the perturbative decay of inflaton in Eq.~(\ref{inflaton}), one may multiply the above solution by the decaying factor $e^{-\hat a}$. Next, with Eq.~(\ref{inflaton:sol}), the energy density of radiation background in the phase II can be obtained by directly integrating over $\hat a$,
\begin{align}\label{radi:sol}
R(\hat a)\approx \f{8}{15}\f{\Gamma_\phi M_{\rm Pl}^2H_{end} }{m_\phi^{4}} \hat a_{end}^{3/2}\hat a^{5/2}.
\end{align}
Whereas in the phase III, the solution to Eq.~(\ref{rd:rasol}) is a trivial redshiting, and then in Eq.~(\ref{DG:re}) $R$ takes a constant value $R_{re}\equiv R(\hat a_{re})$.  Finally, the energy density of dark gluons will be presented at the end of this subsection.

We need the relation between $a$ and $T$ to solve the BEs. After the radiation reaching thermal equilibrium at $t_{th}$, the visible sector temperature can be defined through $\rho_R=\f{\pi^2}{30}g^{re}_* T^4$ with $g_{re}$ the degree of freedoms of the relativistic particles during reheating. Along with Eq.~(\ref{radi:sol}), one gets the scaling rule $T\propto a^{-3/8}$ during the reheating era after thermalization:
\begin{eqnarray}\label{Ta:re}
T=T_{th}\hat a_{th}^{3/8}{\hat a}^{-3/8},
\end{eqnarray}
with $T_{th}$ the thermalization temperature, corresponding to the time scale $t_{th}$. It is also the maximum temperature of the thermalized universe, $T_{max}$. From the above scaling rule and Eq.~(\ref{thermalization}), Eq.~(\ref{ta:re}), one can determine
\begin{align}\label{Tmax}
T_{max}=T_{th}\simeq \alpha^{4/5}m_\phi\L\f{24}{\pi^2 g_{re}}\R^{1/4}\L\f{\Gamma_\phi M_{\rm Pl}^2}{m_\phi^3}\R^{2/5}.
\end{align}
So $T_{max}\sim \alpha^{4/5}\L k M_{\rm Pl}/m_\phi\R^{2/5}m_\phi$. Its magnitude relative to $m_\phi$ depends on the value of $k$. The referred value $k\sim k_0 \equiv m_\phi/M_{\rm Pl}$ is of special interest since it makes $T_{re}\ll T_{max}\ll m_\phi$, which implies that the thermal effects from radiation is fully under control. However, in general $T_{max}$ can exceed $m_\phi$ as $k\gg k_0$ and $m_\phi$ in the relatively low mass region. We have to remind the readers that if $T_{max}>M_{6,8}$, again the on-shell production of mediators may be important. Another merit of using $k_0$ is the greatly narrowing the many possibilities in properly reheating the dark gluonic sector. Hereafter we will quote $k_0$ frequently in the following analysis. Maybe $k_0$ relates to the inflaton decay via the Planck scale suppressed $d=5$ operator like $\phi \bar\psi \psi/M_{\rm Pl}$. After the inflaton decays away, entropy is conserved and thus temperature drops much faster, following the well-known rule
\begin{eqnarray}\label{Ta:rd}
T=T_{re}a_{re} a^{-1}.
\end{eqnarray}
With Eq.~(\ref{Ta:re}) and Eq.~(\ref{Ta:rd}) one can express the collision terms in the light of $a$.

\subsubsection{The cross sections}

We now move to the calculation of the collision terms in Eq.~(\ref{darkG}), which involve the cross sections for dark gluons prodcution. In our setup, the dark gluons are produced via the $2\ra 2$ processes. In general, the Lorentze invariant cross section of the process $1+2\ra 3+4$ is defined as~\cite{average}
\begin{eqnarray}\label{}
\sigma(s)= \f{1}{4\sqrt{(p_1\cdot p_2)^2-m_1^2m_2^2} }\int \f{d^3p_3}{(2\pi)^32E_3}
\f{d^3p_4}{(2\pi)^32E_4}(2\pi)^4\delta^4(p_1+p_2-p_3-p_4)|{\cal M}|^2,
\end{eqnarray}
where the spin summation  for the final state and average of the initial states are implied. The prefactor is known as the Moller velocity. In the massless limit, it is $1/(2{s})$. Because $\sigma(s)$ is Lorenz invariant, we will calculate it in the CM frame of the incident partiles for simplicity. From the effective operators given in Eq.~(\ref{EFT}), the squared amplitudes for the two representative cases are given by
\begin{align}\label{}
d=6:&~~~~~~~|{\cal M}|^2=2\times 2(N_d^2-1)g_d^4\f{s^2}{M_6^4},\\
d=8:&~~~~~~~|{\cal M}|^2=2(N_c^2-1)\times 2(N_d^2-1)g_d^4\f{s^4}{M_8^8},
\end{align}
with $N_c=3$ for QCD. The process $H+H^*\ra g_{d}+g_{d}$ happens before electroweak spontaneously breaking, so there is a factor 2 to account for the doublet. Then, the invariant cross sections are
\begin{align}\label{X6}
d=6:&~~~~~~~\sigma(s)=\f{1}{D_H^2}2(N_d^2-1)\f{g_d^4}{16\pi}\f{s}{M_6^4},\\
d=8:&~~~~~~~\sigma(s)=\f{1}{D^2_g}(N_c^2-1)(N_d^2-1)\f{g_d^4}{16\pi}\f{s^3}{M_8^8}.\label{X8}
\end{align}
Averages of the initial state over the spin and internal degrees of freedom give rise to the factors $1/D_H^2=1/2^2$ and $1/D_g^2=1/(2^2(N_c^2-1)^2)$ for the $d=6$ and $d=8$ cases, respectively. Additionally, we have taken into account the symmetry factors $1/2$ (for $d=6$) and $1/2^2$ (for $d=8$). To guarantee the validation of the effective theory throughout the universe evolution after inflation, it is required that the heavy states integrated out should have masses $\sim M_{6,8}\gg m_\phi$. Otherwise one should take into account the production of mediator~\cite{Chen:2017kvz}.

\subsubsection{The collision term}

With the Lorenz invariant cross sections Eq.~(\ref{X6}) and Eq.~(\ref{X8}), the collision term in Eq.~(\ref{darkG}) can be calculated using the standard approach~\cite{average}.

Let us start from the phase I, the prethermal phase. We need the distribution function of radiation, $f_R(E,t)$, which can be derived by such a fact: At a certain time $t_i$, the radiation from inflaton decay at rest carries a fixed momentum $m_\phi/2$; due to the expansion rule Eq.~(\ref{ta:re}), this spectrum is universally redshifted to the later time $t$, located at the momentum $(m_\phi/2E)^{3/2}$. Using this argument, the spectrum of the radiation is derived to be~\cite{Garcia:2018wtq,Harigaya:2013vwa}
\begin{eqnarray}\label{}
f_R(E,t)\simeq 24\pi^2\f{n_R}{m_\phi^3}\L\f{m_\phi}{2E}\R^{3/2},
\end{eqnarray}
where $n_R$ is the number density of the radiation, obtained by means of directly counting the number of inflatons that have decayed away (assuming that one radiation is produced for per inflaton decay),
\begin{eqnarray}\label{}
n_R\simeq n_{\phi,end}\L\f{a}{a_{end}}\R^{-3}d_\phi,
\end{eqnarray}
where $d_\phi\equiv\Gamma_\phi(t-t_{end})\ll1$ and $n_{\phi,end}={\rho_{end}}/{m_\phi}$, the inflaton number density at the end of inflation. With the above distribution function, one can calculate the collision term
\begin{eqnarray}\label{}
\wt\gamma_{g_d}&\equiv&\int d\Pi_{R}d\Pi_{R'}E_{R}f_Rf_{R'} \int d\Pi_{g_d}d\Pi_{g'_d}(2\pi)^4\delta^{4}(p_d+p_{d'}+p_R+p_{R'})
|{\cal M}(RR'\ra g_dg_d')|^2\cr
&=&18D_RD_{R'}{n_R^2}{m_\phi}\int_1^\infty dx \left[ \sqrt{x^2-1}-\arcsin\sqrt{1-x^{-2}}+x-1\right]x^{-5}\sigma(x)
\end{eqnarray}
with $x\equiv m_\phi/\sqrt{s}>1$ and $D_R$ the internal degrees of freedom of radiation $R$. The Lorentz invariant cross section  $\sigma(x)\propto x^{-2}$ and $x^{-6}$ for the $d=6$ and $d=8$ operators, respectively. And the corresponding integral of $x$ are $\f{1}{30}+\f{\pi}{192}\approx 0.05$ and $\f{1}{90}+\f{7\pi}{5120}\approx 0.015$, giving rise to the following collision terms
\begin{eqnarray}\label{}
d=6&:&~~~~~~~~~~\wt\gamma^I_{g_d}
=0.9\times\f{g_d^4}{16\pi}{2(N_d^2-1)}
\f{n_{\phi,end}^2m_\phi^3}{M_6^4}
\L\f{\hat a}{\hat a_{end}}\R^{-3}\Gamma_\phi^2H_{end}^{-2},\\
d=8&:&~~~~~~~~~~\wt\gamma^I_{g_d}
=0.27\times\f{g_d^4}{16\pi}(N_c^2-1)(N_d^2-1)
\f{n_{\phi,end}^2m_\phi^7}{M_8^8}\L\f{\hat a}{\hat a_{end}}\R^{-3}\Gamma_\phi^2H_{end}^{-2}
\end{eqnarray}
Stressed again, in the prethermal era there is no conceptual of temperature, and thus the collision term has no dependence on $T$. Moreover, those two kinds of operators demonstrate a common behavior as the scale factor increases, $\propto a^{-3}$. The reason is traced back to the fact that the distribution function is the only source of $a$.

After the thermalization is completed at $t_{th}$, the distribution function takes the well known Bose-Einstein or Fermi-Dirac distributions, but in order to derive a simple analytical expression, both of them are approximated by the Maxiwell-Boltzman distribution. After thermalization, the collision term becomes
\begin{eqnarray}\label{collsion:th}
\wt\gamma_{g_d}=
\f{D_R^2}{16\pi^4}T^7\int_0^{\infty}dz \left[
z^2K_2(z)+e^{-z}(z+1)
\right] z^3 \sigma(z)
\end{eqnarray}
with $z=\sqrt{s}/T$, while $\sigma(z)\propto T^2z^2$ and $T^6z^6$ for the $d=6$ and $d=8$ operators. We would like to pause to make a comment on the upper limit for $z$, which is simply set to infinity in the sense that $\sqrt{s}$ can be way larger than the given $T$. Actually, the dominant integrating range is near $z\sim 10$. This treatment results in no $T$-dependence after integrating with respect to $z$, which just contributes a pure numerical factor, 3074 and $1.77\times 10^7$ for $d=6$ and $d=8$, respectively; consequently, $\wt \gamma_{g_d}\propto T^{2d-3}$. The larger $d$ case benefits from more significant UV enhancement in the $z\sim 10$ region, so it enjoys an impressive numerical enhancement. Note that these big numbers  do not appear in $\wt\gamma_{g_{d}}^I$ in phase I where  temperature does not exist and $\sqrt{s}<m_\phi$. 

Eq.~(\ref{collsion:th}) holds as long as the plasma has been thermalized, both in the reheating and the radiation domination eras, which have different $a-T$ relations Eq.~(\ref{Ta:re}) and Eq.~(\ref{Ta:rd}), thus leading to $\wt\gamma^{II}_{g_d}(\hat a)$ and  $\wt\gamma^{III}_{g_d}(\hat a)$, respectively.  Concretely, in the two phases they are given by
\begin{eqnarray}\label{phII}
d=6&:&~~~~~(\wt\gamma^{II}_{g_d},\wt\gamma^{III}_{g_d})
\approx
\f{12\times 2(N_d^2-1)}{\pi^5}\f{g_d^4}{M_6^4}
\L T_{th}^9\hat a_{th}^{27/8}\hat a^{-\f{27}{8}},
T_{re}^9\hat a_{re}^{9}\hat a^{-9}\R,\\
d=8&:&~~~~~ (\wt\gamma^{II}_{g_d},\wt\gamma^{III}_{g_d})\approx
\f{69120 (N_c^2-1)(N_d^2-1)}{\pi^5}\f{g_d^4}{M_8^8}\L T_{th}^{13}\hat a_{th}^{39/8}\hat a^{-\f{39}{8}},T_{re}^{13}\hat a_{re}^{13}\hat a^{-13}\R.
\end{eqnarray}
The higher dimensional operator leads to a higher negative power of $a$, rather than common.  We are considering the evolution of energy density, so the collision terms gain one more power of $T$, compared with the collision terms in the BEs for number density. This is easily seen by using the $T-a$ relation Eq.~(\ref{Ta:re}): For $d=8$, $\wt\gamma^{III}_{g_d}\propto T^{13}$, whereas it is proportional to $T^{12}$ in Ref.~\cite{Chen:2017kvz}.

\subsubsection{The energy relic density of dark gluons}

With all of those ingredients, we are now at the position to calculate the relic density of dark gluons. The comoving energy density of dark gluons at $\hat a_f>\hat a_{re}$ is obtained by integrating over the scaled scale factor $\hat a$~\footnote{Different than Ref.~\cite{Chen:2017kvz}, here we adopt $a$ instead of $T$ as the integration variable. The reason is understood by nothing but that there is no temperature in the prethermal era.} in the prethermalization, thermalization and radiation dominant era successively until $\hat a_f$, 
\begin{eqnarray}\label{}
G_d(\hat a_f)\approx \f{\sqrt{3}M_{\rm Pl}}{m_\phi^6} \Phi^{-\f{1}{2}} \L\int_{\hat a_{end}}^{\hat a_{th}} \wt\gamma^{I}_{g_d} {\hat a}^{9/2}d{\hat a}+\int^{\hat a_{re}}_{\hat a_{th}} \wt\gamma^{II}_{g_d} {\hat a}^{9/2}d{\hat a}+\sqrt{\f{\Phi}{R_{re}}}\int^{\hat a_f}_{\hat a_{re}}\wt\gamma^{III}_{g_d} \hat a^{5}d{\hat a}\R,
\end{eqnarray}
As a feature of UV-freeze in production, it is soon frozen as the scale factor $\hat a_f$ becomes significantly larger than $\hat a_{re}$. The frozen maybe happen even much earlier, and we will come back to this point later.

To reveal the mechanism of production, we present the contributions phase by phase. The integrals are trivial and has simple but lengthy analytical expressions:
\begin{description}
  \item[Phase I] In the phase I, dark gluons production is ``IR" dominated both for the $d=6$ and $d=8$ operators (or any other $d$ in the more general context), because their collision terms share the same power of $a$, $-3$. Taking $a_{th}\gg a_{end}$, one has
      \begin{align}\label{}
d=6:&~~~~~G^{I,6}_d(\hat a_{th})\approx 0.1\alpha^{-16/3}\f{g_d^4}{\pi}(N_d^2-1)k
\f{M_{\rm Pl}m^{1/3}_\phi H^{8/3}_{end}}{M_6^4}
{\hat a^{4}_{end}},\\
d=8:&~~~~~G^{I,8}_d(\hat a_{th})\approx 0.15(N_c^2-1)\f{m_\phi^4M_6^4}{M_8^8 } G^{I,6}_d(\hat a_{th})
\end{align}
Thereby, increasing the dimension of operators leads to a suppressed production yield in the phase I. This contribution has a strong dependence on $\alpha$, stemming from the dependence on $\hat a_{th}$. But $\alpha$ will not appear in the contributions from other phases.
  \item[Phase II] In the phase II, freeze-in production for the $d=6$ case is again IR dominated, and the contribution is
      \begin{align}\label{}
d=6:~~~~~G^{II,6}_d(\hat a_{re})\approx \f{32678}{17}2^{\f{5}{12}}3^{\f{7}{12}}
\f{N_d^2-1}{\pi^{19/2}}\f{g_d^4}{g_{re}^{9/4}}k^{5/6}
\f{M_{\rm Pl}^{11/3}}{M_6^4m_\phi^{7/3}}H_{end}^{8/3}
  \hat a_{end}^{4},
\end{align}
 where $a_{re}\gg a_{th}$ is assumed. The pure numerical factor including the $\pi$ is $\approx 0.046$. For $d=8$, the integral is, 
     \begin{align}\label{II:68}
d=8:~~~~~G^{II,8}_d(\hat a_{re})\approx \f{235008}{\pi^2}\f{k^2(N_c^2-1)M_6^4m_\phi^4}{g_{re}M_8^8}G^{II,6}_d(\hat a_{re}).
\end{align}
The pure numerical factor is about $ 1.9\times 10^5$ taking $N_c=3$, which is an impressive enhancement. Note that the one more power of $T$ of $\wt \gamma_{g_d}^{II}$ commented below Eq.~(\ref{phII}) is not sufficient to make $G^{II,8}_d$ turn into UV dominance, and thus we have the relation shown in Eq.~(\ref{II:68}).

  \item[Phase III] The contribution from Phase III is always UV dominated. This leads to the same expression of $G^{III}_d$ as in $G^{II}_d$, up to the pure numerical factors: $G^{III,6}_{d}\approx 1.37G^{II,6}_{d}$ and $G^{III,8}_{d}\approx 0.17G^{II,8}_{d}$.
\end{description}
The total comoving energy density of dark gluons is $G^{6/8}_d(\hat a_f)=G_d^{I,6/8}(\hat a_{th})+G_d^{II,6/8}(\hat a_{re})+G_d^{III,6/8}(\hat a_f)$, which as mentioned before is frozen soon after reheating. Therefore, one may take $G^{6/8}_d(\hat a_f)$ as a constant $G^{6/8}_d$ as long as $\hat a_f$ is sufficiently large.

It is of interest to estimate the ratio of the contributions from thermal and nonthermal productions,
 \begin{eqnarray}\label{}
d&=&6:~~~~~\f{G^{II,6}_d+G^{III,6}_d}{G^{I,6}_d}\simeq 1.8\times 10^{-15}k^{-1/6}
\L\f{\alpha}{0.01}\R^{16/3} \L\f{100}{g_{re}}\R^{9/4}
\L\f{ M_{\rm Pl}}{m_\phi}\R^{8/3},\\
d&=&8:~~~~~\f{G^{II,8}_d+G^{III,8}_d}{G^{I,8}_d}\simeq2.2\times 10^{-13} k^{11/6}
\L\f{\alpha}{0.01}\R^{{16/3}} \L\f{100}{g_{re}}\R^{13/4}\L\f{ M_{\rm Pl}}{m_\phi}\R^{8/3},
\end{eqnarray}
It is consistent with the argument at the beginning of this section: The production in the prethermal phase may become dominant for a very heavy $m_\phi$, in particular for the case with a larger $d$, which likely belongs to this situation except for a very light $m_\phi$ even below the weak scale. So, Our conclusion is consistent with the one drawn in Ref.~\cite{Garcia:2018wtq}.

\subsection{DGB Relic density}

\subsubsection{A much cooler dark sector after reheating}

To estimate the relic density of DGB, we adopt the working assumption that the dark gluons are thermalized and establish their own plasma temperature ${\cal T}$ not later than $t_{re}$. Moreover, we approximate that the energy transfer between the dark and visible sectors completely ceased just at $t_{re}$ hence  $G_d({\hat a= \hat a_{re}})\approx G_d({\hat a\gg \hat a_{re}})$, amounting to dropping the part from phase III. The resulting error is negligible if UV freeze-in is dominated by phase I; the error is still insignificant even if phase I is not important, because the contributions from phase II and III are comparable. 
   
Now we are capable of computing the ratio of the temperatures of the two sectors at the reheating time. The temperature of the dark gluon plasma  at this time can be derived via  $\rho_{g_d}=\pi^2(N_d^2-1){\cal T}_{re}^4/15$, which gives 
 \begin{eqnarray}\label{}
{\cal T}_{re}\simeq\L\f{\pi^2}{15(N_d^2-1)}\R^{1/4}G^{1/4}_d( \hat a_{re})a^{-1}_{re}.
\end{eqnarray}
Since after reheating both sectors keep cooling down, respectively following the rules ${\cal T}(a)={\cal T}_{re}a_{re}/a$ and Eq.~(\ref{Ta:rd}), from which we immediately get ${\cal T}=\f{{\cal T}_{re}}{T_{re}}T=    \xi_T  T$. As a natural consequence of our mechanism reheating the dark sector, $\xi_T\ll 1$, which just provides the desired hierarchy to accommodate the correct DGB relic density. Let us show the expression of $\xi_T$ for $d=6$:
\begin{eqnarray}\label{}
\xi_{T,6}\simeq 0.65\f{k^{\f{3}{8}}}{g_{re}^{{5}/{16}}} \f{g_dM_{\rm Pl}}{M_6} \L\f{ m_\phi}{M_{\rm Pl}} \R^{3/4}\left[ 0.45 \f{g_{re}^{9/4}}{N_d^2-1}\alpha^{-16/3}k^{\f{1}{6}}\L\f{ m_\phi}{M_{\rm Pl}} \R^{8/3}+ 1\right]^{\f{1}{4}}.
\end{eqnarray}
Increasing $m_\phi$ helps to enhance $\xi_T$. The first and second terms in the square bracket originate from phase I and phase II plus phase III, respectively. To have a more direct impression on this ratio, let us analyze two limits. 
\begin{itemize}
  \item For a relatively light $m_\phi\ll 10^{13}{\rm GeV}(1.0/k)^{1/16} \L 100/g_{re}\R^{\f{27}{32}}\L\alpha/0.01\R^{2} $, the phase I contribution is absolutely negligible. Note that this condition is insensitive to $k$ because of the power suppression $k^{1/16}$. Now one has the estimation
\begin{align}\label{}
\xi_{T,6}\sim k^{3/8} \f{M_{\rm Pl}}{M_6}\L\f{ m_\phi}{M_{\rm Pl}}\R^{3/4}=r_{\phi,6}k^{3/8}  \L\f{M_{\rm Pl}}{ m_\phi}\R^{1/4}\gg r_{\phi,6}k^{3/8}/ \alpha^{1/2},\nonumber
\end{align}
with $r_{\phi,6}\equiv m_\phi/M_6\ll1$.~\footnote{As we have commented before, $M_{6}$ is not exactly corresponding to the mass scale of loop particles, but in the concrete UV model the actual mass scale usually is even lower than $M_6$, except for the presence of large couplings in the loop which could overcome the loop suppression.} The smallness of $\xi_{T,6}$ is readily obtained by a sufficiently small $r_{\phi,6}$ or $k$. For the reference value $k=k_0$, $\xi_{T,6}\sim r_{\phi,6}\L{ m_\phi}/{M_{\rm Pl}}\R^{1/8}$ and then how to get a large enough $\xi_{T,6}\gtrsim 10^{-3}$ is of concern; hence instead $r_{\phi,6}$ should have a moderate size.
  \item To the contrary, if $m_\phi\gg 10^{13}{\rm GeV}$, then the phase I is the dominant contribution to reheat the dark sector and $\xi_{T,6}\sim k^{5/12} \alpha^{-\f{4}{3}}r_{\phi,6} \L\f{ m_\phi}{M_{\rm Pl}}\R^{5/12}$. 
On the other hand, in cosmology the inflation models  having convex potentials give the largest inflaton mass, e.g., in the popular chaotic model $m_\phi\lesssim 10^{15}$ GeV~\cite{chaotic}. This means that there is barely room for this scenario. It is tempting to fix $m_\phi=10^{15}$ GeV for this case, dubbed the convex limit. But will not expand discussions on this special case.
 
\end{itemize}

A similar analysis can be made for the case with $d=8$, yielding
\begin{eqnarray}\label{}
\xi_{T,8}\simeq 0.55g_dk^{\f{5}{12}}
\alpha^{-\f{4}{3}}g_{re}^{\f{1}{4}} r_\phi^2\L\f{ m_\phi}{M_{\rm Pl}} \R^{-\f{1}{4}} \left[ \L\f{ m_\phi}{M_{\rm Pl}} \R^{8/3}+ \f{174932}{g_{re}^{{13}/{4}}}\alpha^{\f{16}{3}}k^{\f{11}{6}}\right]^{\f{1}{4}},
\end{eqnarray}
where $N_c=3$ has been used. The first term denotes for the contribution from phase I, and it dominates the other contributions when
\begin{eqnarray}\label{d8Ido}
m_\phi\gg 0.4\times 10^{13}{\rm GeV}\L{\alpha}/0.01\R^{2} \L 100/g_{re}\R^{39/32}\L k/1.0\R^{11/16},\nonumber
\end{eqnarray}
almost coinciding with the counterpart in the $d=6$ case. But here taking a very small $k$ could substantially relax the condition. For instance, consider $k=k_0$ then the above condition is weakened to $m_\phi\gtrsim 10^{3}$ GeV, which is a loose condition. Compared to the $d=6$ case, where it is hard to realize the phase I production of dark gluons, in the $d=8$ case this phase tends to play a much more prominent role, in particular for a very slowly decaying inflaton (namely $k\ll 1$). The cause is traced back to the enhanced UV-sensitive by virtue of the larger $d$. Choosing $k=k_0$ has one more reason, though not physics related: The discussions on prethermalization production of dark species is rare, and here we present one good example. Anyway, in the $d=8$ case we will just focus on the scenario of prethermal production dark gluons and then $\xi_{T,6}$ is estimated to be
\begin{eqnarray}\label{}
\xi_{T,8}\simeq 8.1g_d\L0.01/{\alpha}\R^{4/3} \L g_{re}/100\R^{1/4}\L r_\phi/0.1\R^2\L k{ m_\phi}/{M_{\rm Pl}} \R^{\f{5}{12}}. \nonumber
\end{eqnarray}
Taking $k=k_0$, to make $\xi_{T,8}\gtrsim10^{-4}$ the inflaton must be as heavy as around $10^{13}$GeV. But if we work in the intermediate region $1\gg k\gg k_0$, the required $m_\phi$ can be accordingly lighter.


\subsubsection{Dark confining phase transition and relic density of the dark gluon ball}

As ${\cal T}$ drops below $\Ld$, dark confining phase transition happens, and the relativistic dark gluons are confined to dark glue balls. The energy density of dark gluons is assumed to be transferred to that of the lowest state of DGB, $s$,
\begin{eqnarray}\label{}
\rho_s(a_\Ld)\approx\rho_{g_d}({\cal T}=\Ld_d)=\f{\pi^2}{15}(N_d^2-1)\Ld_d^4.
\end{eqnarray}
This is a reasonable assumption to estimate the DGB relic density. But the excited states may be also produced and moreover the DGBs still carry some momentum (the corresponding kinetic energy does not contribute to the final DM relic density) just after the phase transition, although they soon become nonrelativistic. Estimation of the resulting uncertainty is beyond the scope of this article.

After the confining phase transition, the energy density of the nonrelativistic DGB scales as matter $\propto a^{-3}$ until today, when the visible sector temperature $T_0=2.37\times 10^{-13}$ GeV. Hence, its fraction in the total energy budget $\Omega_sh^2\equiv\f{\rho_s(a_0)}{\rho_c}=\f{\rho_s(a_\Ld)}{\rho_c}
\L\f{a_\Ld}{a_0}\R^{3}$ is
\begin{eqnarray}\label{relic}
\Omega_sh^2=\f{\rho_S(a_\Ld)}{\rho_c}
\L\f{\xi_TT_0}{\Ld_d}\R^{3}=0.16\times (N_d^2-1)\f{\pi^2}{15}\L\f{\xi_T}{0.001}\R^3\L\f{\Ld_d}{\rm GeV}\R,
\end{eqnarray}
where we have used the critical energy density $\rho_c=8.1\times 10^{-47}h^2{\rm GeV}^4$. Note that here we estimate the DGB relic density directly from energy conservation, but the precise relic density of DGB relies on the $s^5$ term in the low energy effective model of DGB, namely Eq.~(\ref{EFT}). The two approaches give the same scaling behavior $\Omega_sh^2\propto \xi_T^3\Ld_d$. 
\begin{figure}
\includegraphics[width=1.55in]{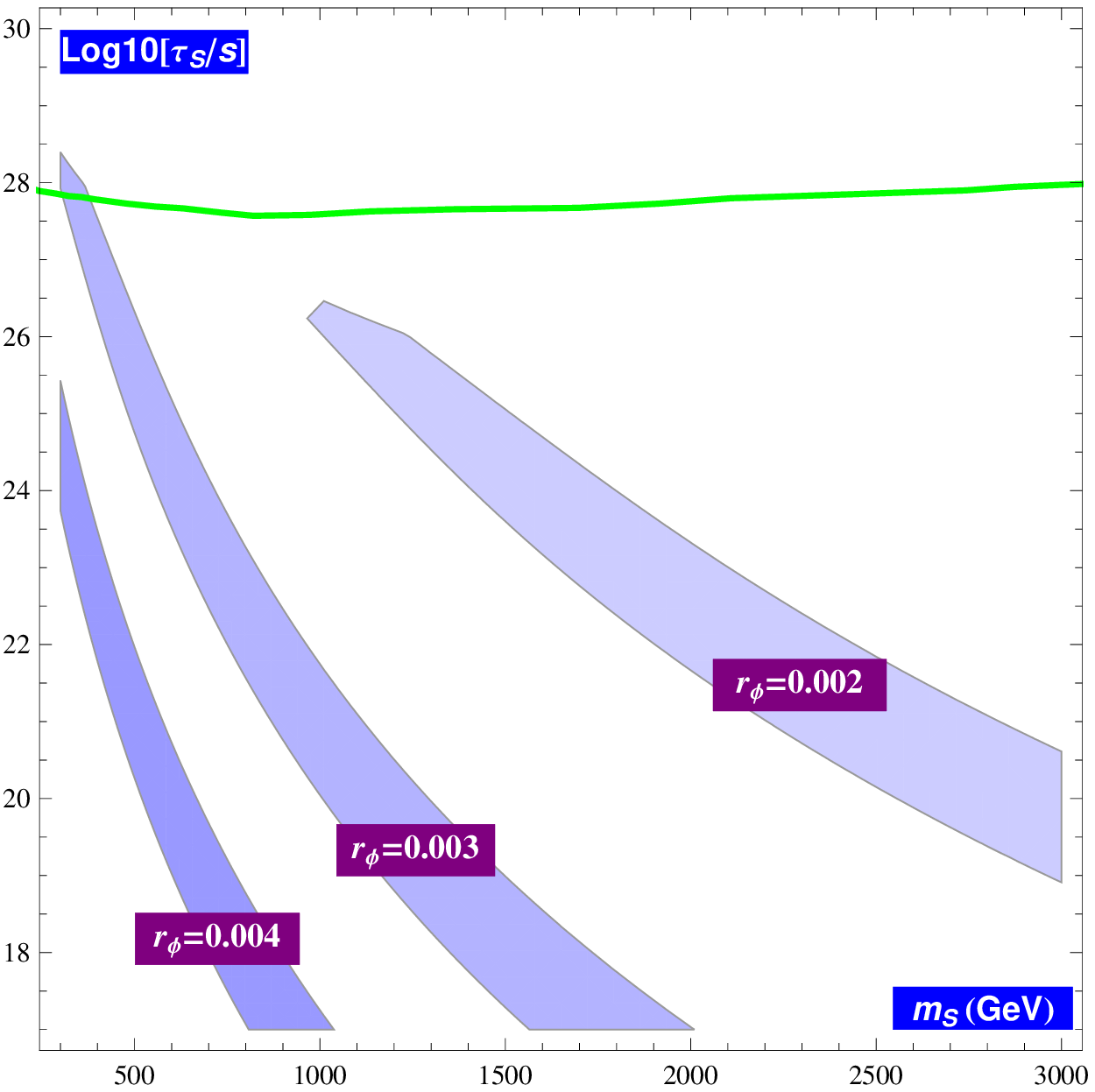}
\includegraphics[width=1.55in]{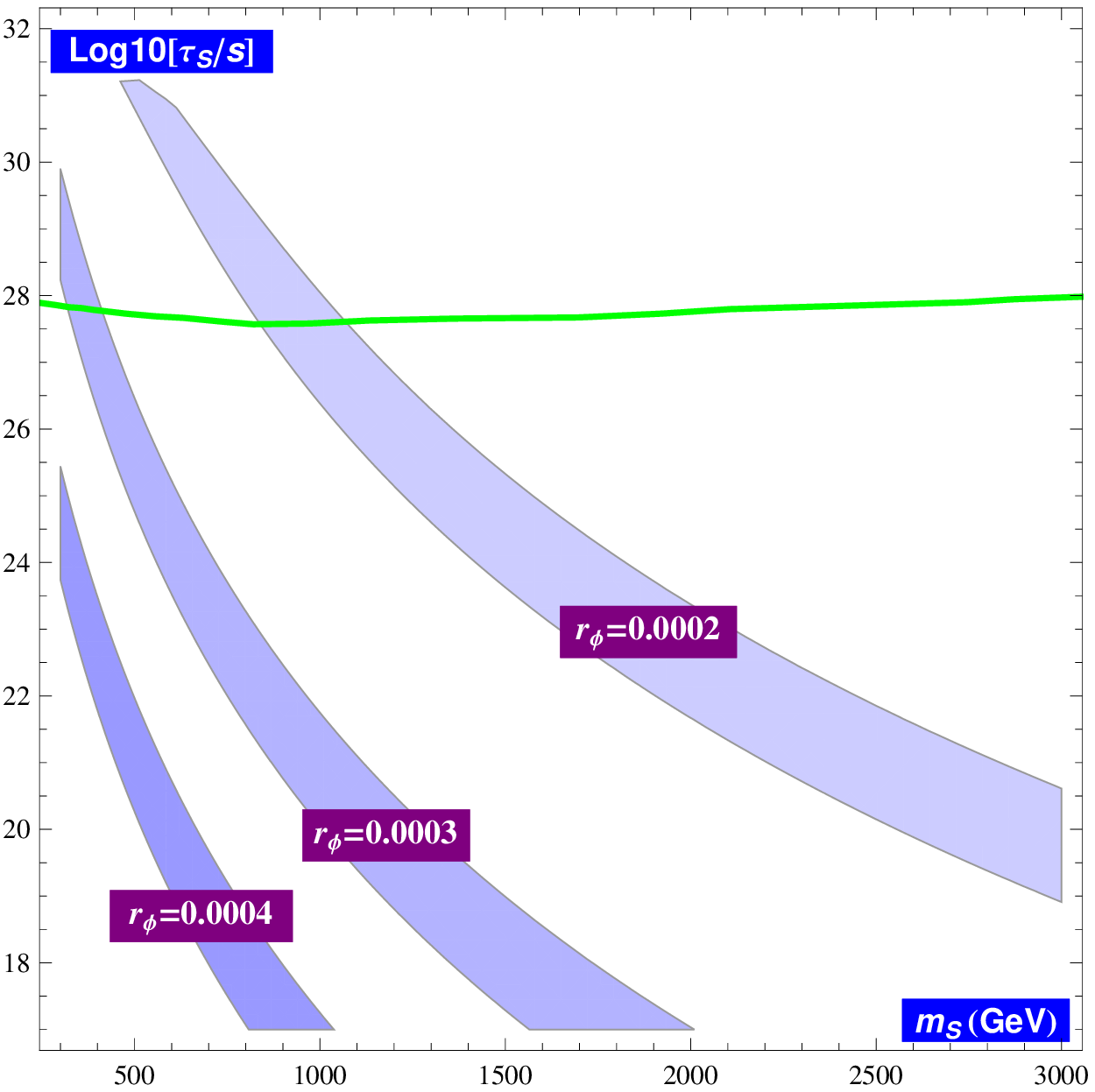}
\includegraphics[width=1.55in]{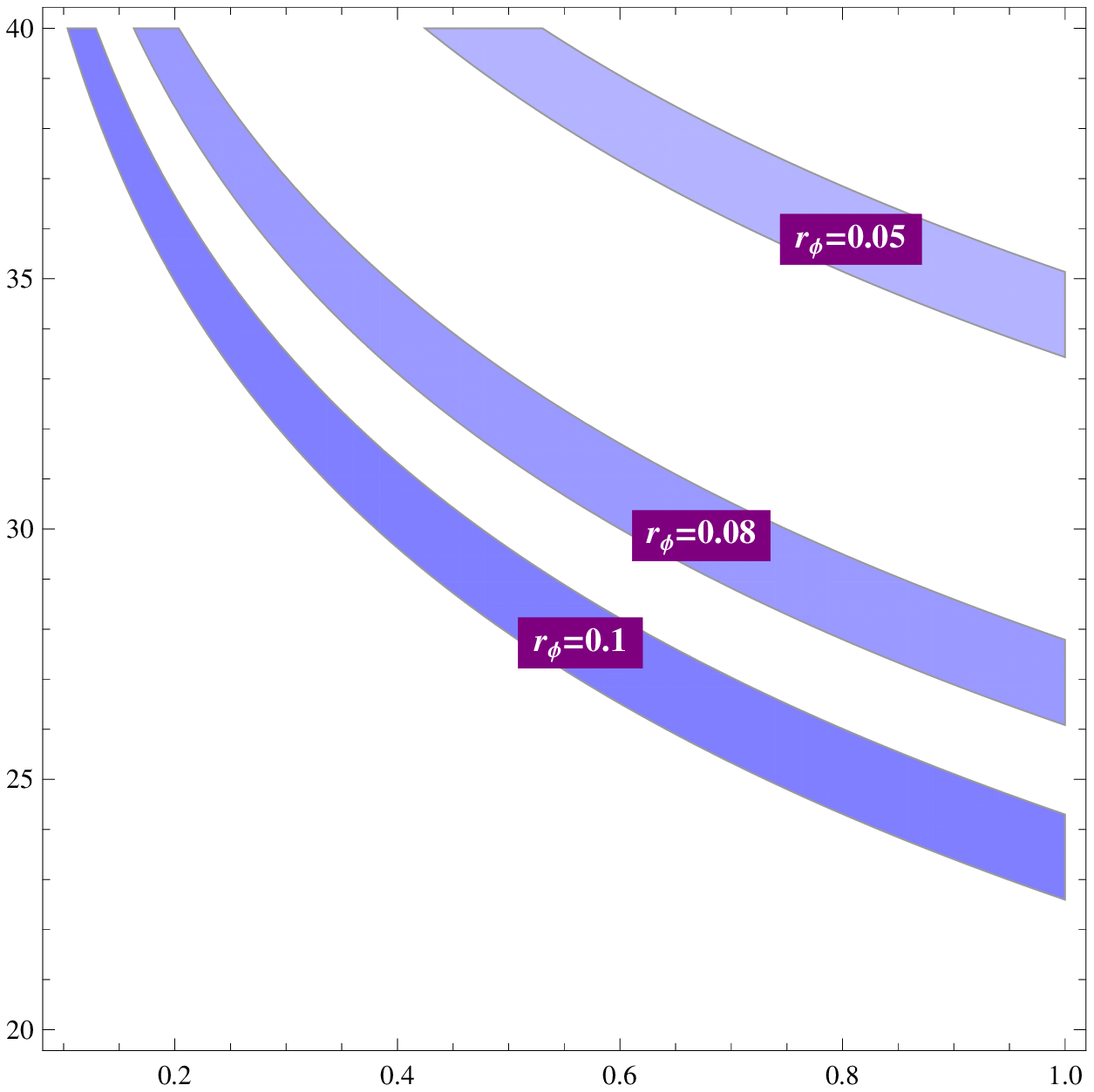}
\includegraphics[width=1.55in]{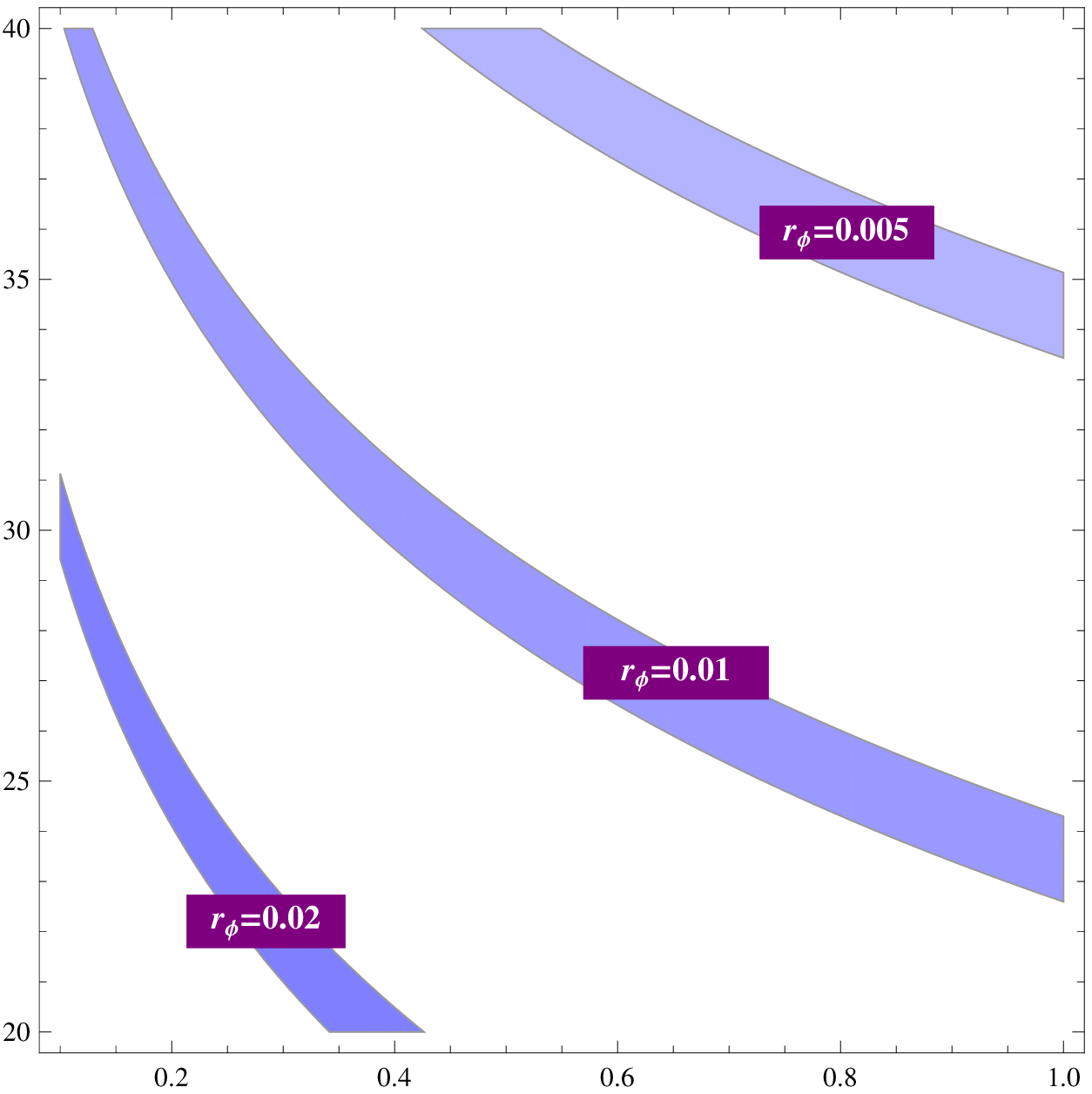}
\includegraphics[width=1.56in]{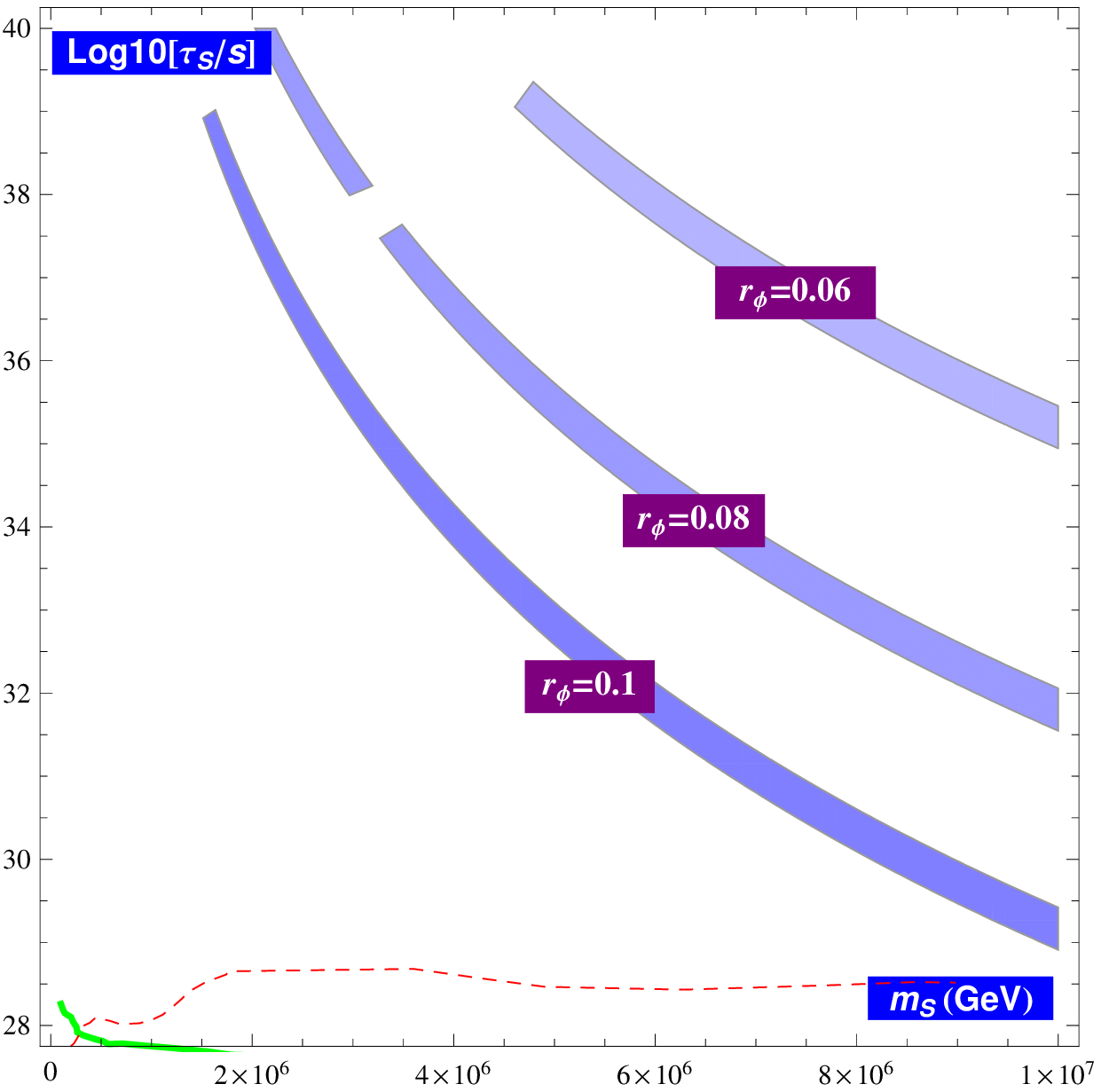}
\includegraphics[width=1.56in]{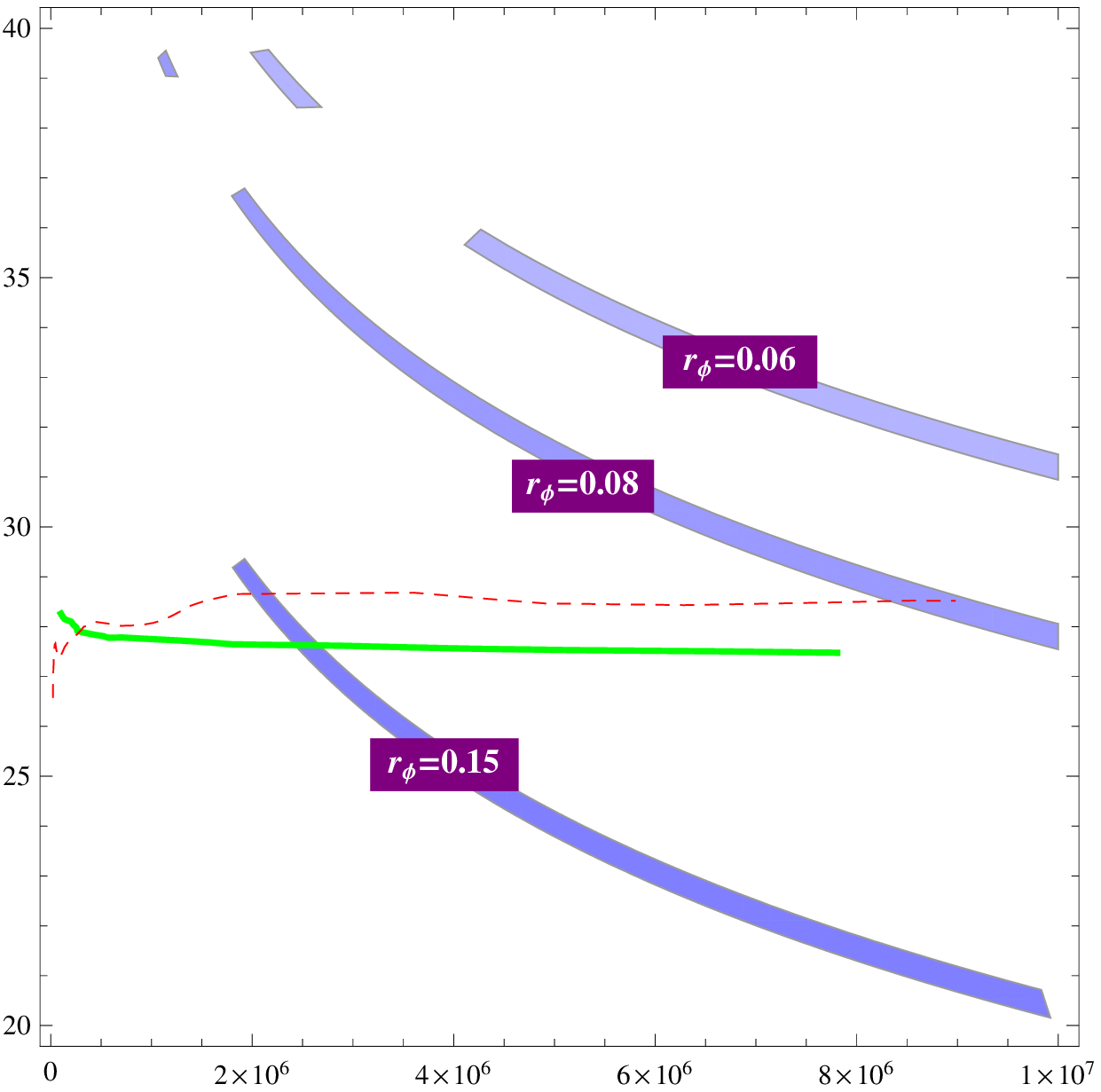}
\includegraphics[width=1.56in]{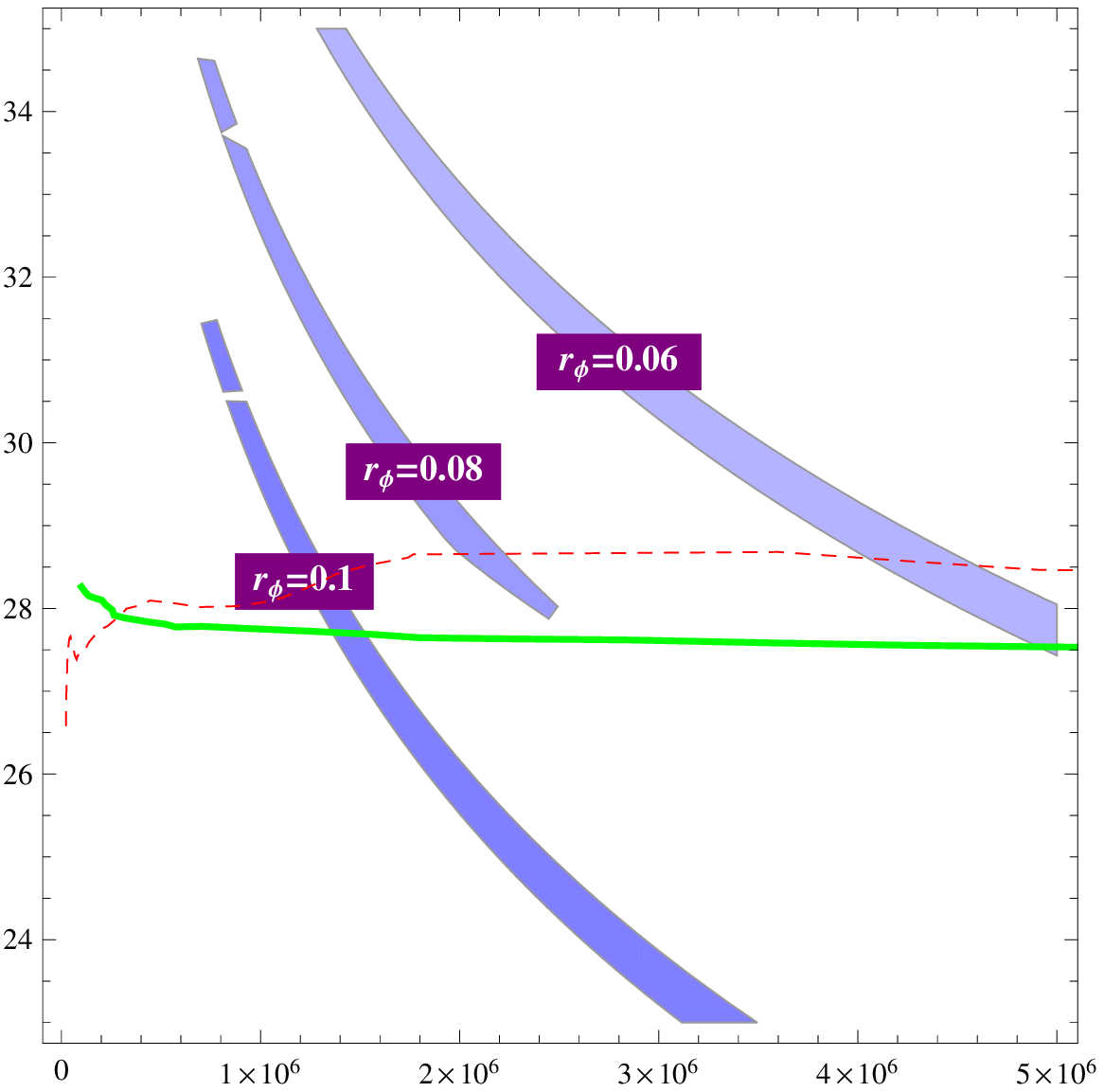}
\includegraphics[width=1.56in]{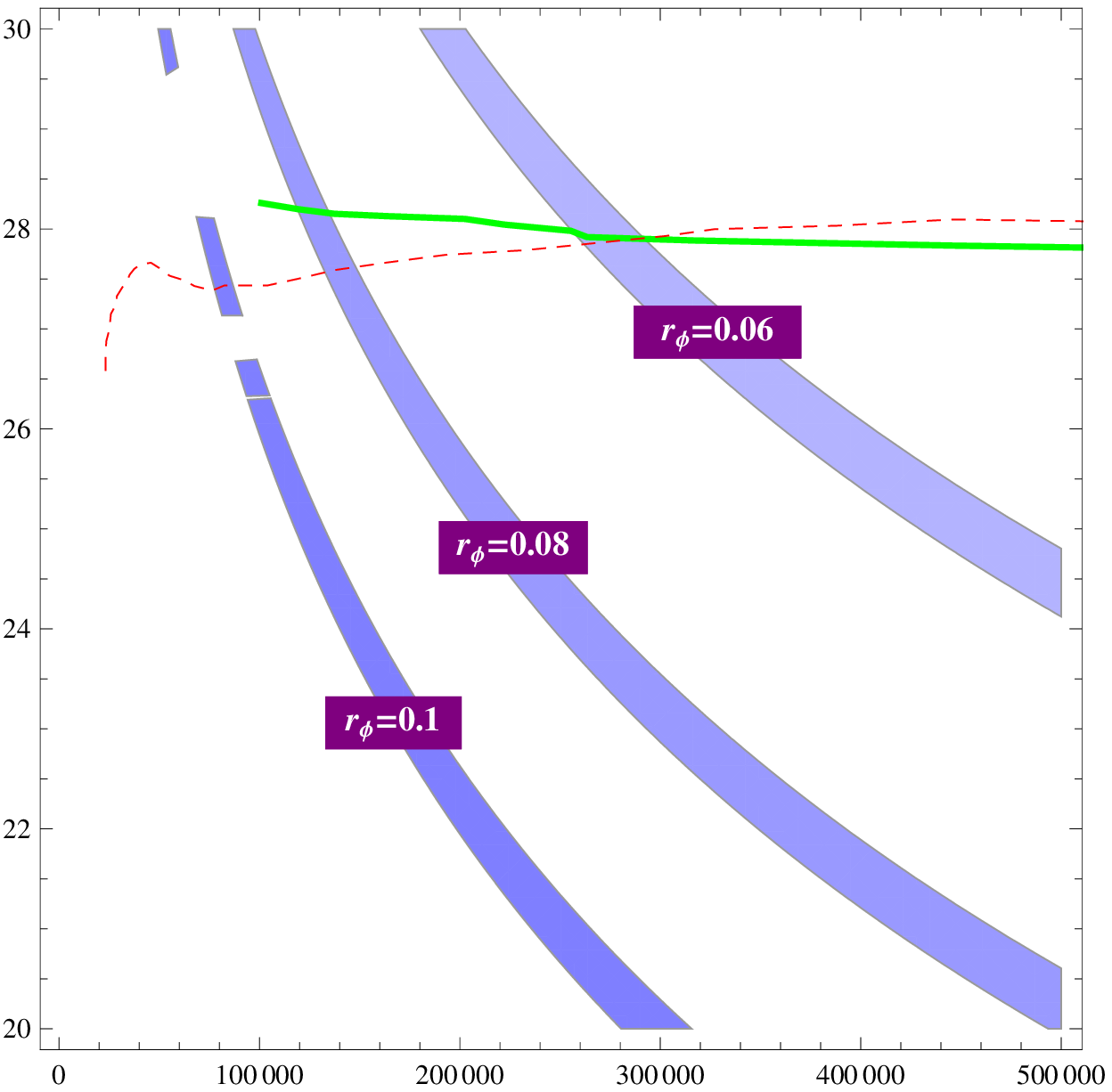}
\caption{DGB dark matter shown in the $m_{s}-\tau_{s}$ plane for $d=6$ (upper) and $d=8$ (lower), taking $N_d=5$. We choose several $r_\phi$ values labeled in purple; we allow a small deviation of DGB relic density from the exactly correct value, $0.13>\Omega_sh^2>0.09$, which leads to a band instead of a line. We also require $m_\phi<10^{15}$GeV. Upper panels: The first ($k=k_0$) and second ($k=1000k_0$)panels are for the relatively heavier DGB much above the weak scale which dominantly decays into a pair of vector bosons, while the third ($k=k_0$) and fourth panel are for the relatively light DGB which dominantly decays into $\mu\bar \mu$. Lower panels: first ($k=k_0$) and second ($k=10k_0$) for prethermal UV freeze-in; third ($k=10^3k_0$) and fourth ($k=10^4k_0$) for thermal UV freeze-in. The thick green line and dahsed red line denote the FERMI-LAT and IceCube constraints, respectively.}\label{d=6}
\end{figure}

As we have stressed, a small $\xi_T$ is a built-in feature of our way to reheat the dark sector and thus the relic density problem can be naturally solved. It is still of concern that if the DGB could leave observable via its decay. To that end, we demonstrate the parameter space of DGB with correct relic density on the $(m_{s}, \tau_{s})$ plane in Fig.~\ref{d=6}, choosing several values of $r_\phi$. This is done by trading $M_{6,8}$ with $\tau_{s}$ in Eq.~(\ref{SVV}) or Eq.~(\ref{Smu}), $\Ld_d$ with $\Omega_sh^2$ via  Eq.~(\ref{relic}). Several observations are made:
\begin{itemize} 
  \item In general, for a given DGB mass, a larger $r_\phi$ leads to a significantly shorter DGB lifetime. This is because, to maintain a constant $\xi_T$, accordingly $M_{6,8}= m_\phi/r_\phi$ becomes smaller thus much faster DGB decay. In other words, the parameter space having a larger $r_\phi$ demonstrate a more promising detect prospect. Partial parameter space which gives DGB lifetime significantly shorter than $10^{28}s$ has already been ruled out. We quote the results from Ref.~\cite{Cohen:2016uyg}, displaying the FERMI-LAT gamma ray data~\cite{FERMILAT} and IceCube neutrino data~\cite{IceCube} constraints in thick green and dashed red lines.
  \item 
In the $d=6$ namely Higgs portal case (upper panels), as explained before, it is hard to realize the scenario of dark gluon production mainly in the phase I, so all four panels ${(G^{II,8}_d+G^{III,8}_d)}/{G^{I,8}_d}>5$. In the heavy DGB region $\sim {\cal O}(\rm TeV)$, referring to the first and second panels, it is clearly seen that the larger $r_\phi$ cases, in particular given a larger $k$, have been ruled out by FERMI-LAT already. Whereas the low mass region is difficult to probe, owing to both its soft final states and much longer lifetime. 
\item
In the $d=8$ case (lower panels), the decay rate is highly suppressed by $(m_s/M_8)^9$, while this suppression does not appear in the relic density where $m_s$ is replaced by the much higher scale $m_\phi$, so this case stands little chance of observation in the light DGB region. To get a larger decay rates, we merely focus on the region above PeV scale; increasing $k$ could help to enhance the decay rates and thus some of the parameter space is accessible in the experiments which are sensitive to super energetic cosmic ray, for instance the gamma ray data (up tp 2TeV) from FERMI-LAT and high energy neutrino data from IceCube. We are considering the gluon-portal and its contributions to gamma ray and neutrinos require simulations. But if one replaces gluons with the electroweak gauge bosons, then one may interpret the data similar to the Higgs portal, to find that it is sensitive to lifetime $\gtrsim 10^{28}s$ for a PeV scale DGB~\cite{Cohen:2016uyg}. 
\end{itemize}
Anyway, in this paper we just schematically show the features of the parameter space which may admit a future discovery, and the systematic discussions on the indirect detection bounds deserves a specific study elsewhere.

\section{Conclusion and discussions}

Many new physics give rise to a pure $SU(N_d)$ gauge sector surviving at low energy, and dark glueball is a prediction of such a sector in the confining phase. DGB provides a simple non-WIMP DM candidate characterized by very few parameters. However, its correct relic density needs careful treatment because it in general leaves too many relics to overcome the universe. The way out this problem is  that the gluonic sector should be much cooler than the visible sector after the two sectors are reheated. We pointed out that if the two sectors are linked via the higher dimensional operators, and the SM species freezes in the gluonic sector in the very early universe, a very cool gluonic sector is a natural consequence. For concreteness, two kinds operators are introduced:
\begin{itemize}
  \item At the $d=6$ level the dark gauge field strength tensor couples to the SM Higgs doublet; 
  \item At the $d=8$ level the dark gauge field strength tensor couples to the SM vector bosons.
\end{itemize}
We carefully studied the yields of dark gluons in the different phases of the universe, from the end of inflation, reheating in the prethermalization and thermalization stages to radiation dominance, to find that the temperature and the production of the dark gluonic sector is sensitive to the inflaton mass and decay width; they determine two key temperatures $T_{max}$ and $T_{re}$. For instance, considering the well-motivated Planck suppressed decay of the inflaton, it is found it is difficult to sufficiently reheat the dark gluonic sector; moreover, it is difficult to realize prethermalization production of dark gluons (it is true even for the much faster inflaton decay). By contrast, in the $d=8$ case prethermalization production tends to be the main production mechanism. Anyway, the DGB is a viable dark matter candidate over a wide mass region, from sub-GeV to multi-PeV. The higher dimensional operators at the same time open decay channels for the DGB, and some of the parameter space leave hints in the cosmic ray: The $d=6$ case is hopeful to be detected in the TeV DGB mass region, while in the $d=8$ case a PeV scale DGB with lifetime $\sim 10^{28}s$ might be probed by FERMI-LAT and IceCube. Interestingly, the IceCube PeV events can be interpreted by a decaying DGB with electroweak gauge bosons portal at the $d=8$ level.

There are some open questions. In this paper we are confined to the assumption that the messengers are very heavy, having mass much greater than $m_\phi$, and thus they can not be produced on-shell during reheat. But this is questionable if one has the interest in a low cut-off scale and thus the messengers can be produced even if the reheat temperature is low. In this case we have to carefully take into account the production of messengers, to see if their own could be thermalized. Another open question is the thermalization of the gluonic sector. If it never establishes thermal equilibrium, there will be no confining phase transition and the dark gluons will leave as dark radition. Actually, thermalization of a generic FIMP sector is also of interest because it may has cosmological implications. Last but not least, a detailed study on the indirect detection bounds in the wide DBG mass region should be done in the coming paper. The possible gravitationals wave associated with the confining phase transition may also furnish a window to probe this simple but fairly hidden DM candidate.

\noindent {\bf{Acknowledgements}}

I would like to thank Institute of Theoretical Physics, Chinese Academy of Sciences for the hospitality during my visiting there to finalize this manuscript. This work is supported in part by the National Science Foundation of China (11775086).





\end{document}